# ARCHITECTURE OF AN OPEN-SOURCED, EXTENSIBLE DATA WAREHOUSE BUILDER: INTERBASE 6 DATA WAREHOUSE BUILDER (IB-DWB)

Maurice HT Ling [1], Chi Wai So [2]

[1] Dept. of Zoology, The University of Melbourne
[2] National Computing Centre, UK (Student)

ABSTRACT: We report the development of an open-sourced data warehouse builder, InterBase Data Warehouse Builder (IB-DWB), based on Borland InterBase 6 Open Edition Database Server. InterBase 6 is used for its low maintenance and small footprint. IB-DWB is designed modularly and consists of 5 main components, Data Plug Platform, Discoverer Platform, Multi-Dimensional Cube Builder, and Query Supporter, bounded together by a Kernel. It is also an extensible system, made possible by the Data Plug Platform and the Discoverer Platform. Currently, extensions are only possible via dynamic linked-libraries (DLLs). Multi-Dimensional Cube Builder represents a basal mean of data aggregation. The architectural philosophy of IB-DWB centers around providing a base platform that is extensible, which is functionally supported by expansion modules. IB-DWB is currently being hosted by sourceforge.net (Project Unix Name: ib-dwb), licensed under GNU General Public License, Version 2.

## OVERVIEW

InterBase Data Warehouse Builder (IB-DWB) is an open-sourced, extensible data mart / warehouse building tool that is developed as an add-on to Borland InteBase 6 Database Server Open Edition. InterBase 6 is chosen as the server due to three reasons. Firstly, it is open-sourced, which allows for code-level optimization and binding, if necessary. Secondly, it has a small installation footprint. Complete installation requires less than 10MB [Tod01]. Lastly, it requires low maintenance. With query optimization [Car01], amongst other tools, performance tuning and maintenance is kept to the minimum. In line with the open-source culture pioneered by InterBase, IB-DWB, as presented here, is an open-sourced software, hosted by Sourceforge.net (and is undergoing further development) (Project Unix Name: ib-dwb).

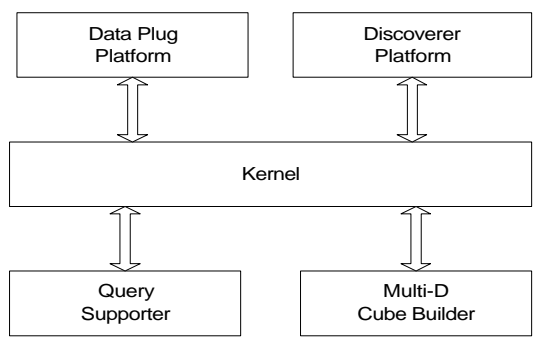

Figure 1: Semantics of IB-DWB

IB-DWB, designed as a platform for incremental and continual expansion, consists of four components, tied together by a kernel component. The four components are, Data Plug Platform, Discoverer Platform, Multi-Dimensional (Multi-D) Cube Builder, and Query Supporter. These components function independently of each other but are linked as shown in Figure 1.

The rest of this paper is organized as follows: a list of the major design philosophy of IB-DWB prior to a brief illustration of some commercial data warehouse builders, before describing each of the five components followed by a before general deployment of the tool. Finally, we will comment on the future developments of this software, and in appendix, segments of specifications are shown to illustrate the individual components.

## RESEARCH ON COMMERCIAL SYSTEMS

In this project, two commercial data warehouses builders, Oracle Warehouse Builder (OWB) and Sybase Industry Warehouse Studio (IWS), were examined in depth, amongst others, including Microsoft SQL

Server 2000 Data Analysis Suite [Pau01]. Microsoft did not have a separate tool for data warehousing, instead, it is available as add-ons to the database server.

OWB consists of 7 components, namely, GUI, a repository which holds a set of internal tables, code generator to generate codes to create and administer the warehouse, integrator for source data extraction, bridge to facilitate data exchange, browser and runtimes [Gio00]. Ad hoc queries and analysis are supported by Oracle9iAS Discoverer, linked to OWB via a bridge [Ora01].

Contrasting to OWB, which assumes nothingness, IWS assumes a 4-layered concentric data model [War99]. In its core is a set of tables and views common to all data warehouses. The second layer (vertical model) comprises of user-defined tables, which allows for industry- and application-specificity. The third layer fine-tunes the first two layers by a set of aggregates and controls. The last layer (data element) holds the actual data.

By studying commercial tools, a few points are observed. Firstly, data warehousing is an extension to traditional databases. Secondly, there need not be a base data model. Thirdly, aggregation is a critical component for data warehouse functionality. Lastly, analysis components can be available as separate modules, as in OWB and Oracle9iAS Discoverer.

MAIN ARCHITECTURAL PHILOSOPHY

IB-DWB is designed with 3 major thoughts in mind. Firstly, it should have a small installation size. Secondly, to allow for virtually limitless expansion, the system should place minimal restriction on expansion modules. Lastly, to minimize program size and ability to be customized to specific needs, only the base functions are to be built-in. With these, IB-DWB, being a base system, consisting of only the core of what is necessary in a data warehouse builder, with nearly all functionalities to be added on by additional modules. Both Apache and Linux are examples of such an assembly.

KERNEL

The kernel component of IB-DWB has a number of important functions. Firstly, it enables data transfer from one component to another. Secondly, it facilitates communication with InterBase 6 server. Finally, it acts as a manager for other components. Being so, it is made up of four components, InterBase Access Plug (IAP), Tool Controller, Wizard Controller, and Kernel Binder. A semantic diagram of the kernel is given in Figure 2.

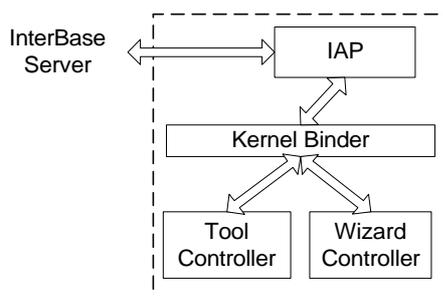

Figure 2: Semantics of Kernel (Kernel is boxed by broken lines)

IAP forms the actual communication channel between InterBase server and IB-DWB. Implementationally, it is a data module comprises of InterBaseExpress components [Ove01] [Tod00]. To cater for functionality, IAP is capable of 5 simultaneous transactions in a single instance of the program. Tools controller and Wizard controller are identical in functionality to Data Plug Platform and Discoverer Platform (see below). They form the kernel level controller for expansion modules. The purpose of redundancy is to act as a backup, as re-installing tools and wizards can take a great deal of effort and time. Kernel binder is the primary layer for communications among the components of the kernel. Effectively, it can be visualized as a cabling line for communication, with no other functions in the current version.

DATA PLUG PLATFORM AND DISCOVERER PLATFORM

We define data plug as a module, which integrates databases or tables in other formats into InterBase database format (GDB). In simple terms, a data plug module is a database converter. A discoverer module is defined as one which is capable of data analysis. Hence, data mining modules are categorized as discoverer modules for the purpose of this tool. As the term, "platform", implies, the data plug platform and discoverer platform are bases of which data plug modules and discoverer modules can be built and linked. These platforms or components form the extensibility of IB-DWB.

Expansion modules can be classified as either tool module or wizard module, the difference being that wizard modules will have initialization string while tool modules do not have. All modules are programmed as dynamic linked-libraries (DLLs) and each set will contain an initialization file containing required installation data. The requirements of the modules are given in Table 1 and the file format of the initialization file is given in Figure 3. The platforms have the role of managing and activating the modules as required. It provides a user interface for management and access to inner controls, such as, tools controller and wizards controller, which are part of the kernel.

Table 1: Module Requirements

| Format | Dynamic linked-libraries |
|---|---|
| Entry | For tool, "TOOLMAIN" function<br>For wizard, "WIZARDMAIN" function |
| Others | One initialization,<br>Return point from DLL<br>User Interfaces (optional) |

```
[SETTINGS]
TYPE=tool
NAME=APriori Text
VERSION=1
AUTHOR=Keith
INIT=pagesize 4096
DESC=a tool using apriori algorithm
```
Figure 3: Initialization File Format

Due to exceeding similarities of both platforms, they are programmatically identical. Hence, their difference boils down to management differences instead of logical differences. Each platform consists of three modules, installation (for installing new modules), un-installation (for removing installed modules) and activation (for executing installed modules) modules.

Each module will consist of at least a DLL file (the actual module) and a text file (initialization file) of the same file name as the DLL file. During installation, the respective platforms will read the initialization file to register the corresponding DLL module. The opposite process explicit removal of a module. Hence, a module can have more than one DLL files, only that the function entry point must be in the DLL with identical file name (excluding extension) to the initialization file. During installation, a check will be made to prevent installation of an earlier version or duplicate installation.

When the installed module is executed, program control is passed to the DLL until it returns. Hence, it will be for the module to request for data from the database, process it, and presents its output. It is deliberate that during such module execution, IB-DWB main system, except InterBase 6 server, plays no role in it. This architecture reduces placed on the module by the system to its minimum.

MULTI-DIMENSIONAL CUBE BUILDER

Multi-Dimensional Cube Builder (Multi-D Cube Builder) is a rudimentary in-built data aggregator in the system. One of its main features is that it does not perform any data manipulations in the original databases nor duplicate the original data for aggregation. Instead, it performs "aggregation *in situ*" by recording only the aggregate mapping. The only restriction imposed is that the same dimension cannot be used twice. This will meant that the largest possible dimension in a cube is the sum of the total number of fields in all tables in the desired database(s). There are no theoretical limits to the numbers of database that can be aggregated as long as it existing.

```
CREATE TABLE CUBETABLE(                     CREATE TABLE DIMENSIONLIST(
   CUBENAME varchar(255) NOT NULL,             CUBENAME varchar(255) NOT NULL,
   CUBEOWNER varchar(255) NOT NULL,            DATABASENAME varchar(255) NOTNULL,
   CUBEDESC varchar(255) NOT NULL,             TABLENAME varchar(255) NOT NULL,
   PRIMARY KEY (CUBENAME));                    DNUMBER integer NOT NULL,
                                               DIMENSION varchar(255) NOT NULL);
```
Figure 4: Definition of CubeTable and DimensionList

"Aggregation *in situ*" is done using two tables, CubeTable and DimensionList. CubeTablet stores the information that is useful to users, such as, description of the cube, while DimensionList holds the information for dimensional aggregation of the cube. The tables are related in a one (CubeTable) to many (DimensionList) fashion. This means is preferred over having DimensionList as an array field in CubeTable, solely for the sake of the ability to assemble extremely high dimensional aggregates, if the need arises.

QUERY SUPPORTER

The main aim of the query supporter is to provide a basic querying capability to the system. There are three reasons to why it is minimalistic in design. Firstly, there is a wide variety of data analysis means and to include even a small subset of it will increase deployment size. Secondly, complex data analysis and query has the tendency to be application-specific. Lastly, by using only the functionalities provided by InterBaseExpress components, it becomes a useful diagnostic tool for both fundamental analysis and initial data exploration.

Essentially, query supporter is little more than a SQL92 select statement interpreter using the SelectSQL method of IBDataSet component in InterBaseExpress.

IMPLEMENTATION

Ease of deployment is a concern during the development of IB-DWB. In this section, we will exemplify a scenario of deployment to illustrate the simplicity of the process.

Currently, "installation" process is in its simplest form. Literally, just deploy the contents of the zipped file and run IBDWB.exe. There is virtually no installation required as no files are registered with Windows registry. However, IB-DWB makes assumptions that InterBase 6 is in "C:\Interbase" and IB-DWB is installed into "C:\IBDWB". Of course, prior to running IB-DWB, InterBase 6 must be active. The next step of setting up is to install the necessary additional modules via Data Plug Platform or Discoverer Platform. Lastly, if the raw data are not in InterBase database format, conversion is necessary. As one might have expected, module installation in either platforms is essentially a registering process, it is possible to pre-customize a set of modules, not unsimilar to a distribution in the context of Linux.

FUTURE DEVELOPMENTS

A beta version of IB-DWB, version 1.0, is available for download and evaluation at sourceforge.net. Current planned developments include developing add-on modules for data analysis, such as, sliding window [Lee01] and M-Apriori [Hol99]. At the same time, Query Supporter will undergo re-vamp to support complex queries of cubes in a GUI.

ACKNOWLEDGEMENT

We will like to acknowledge the work of Lukas Chan, who had contributed during the starting phase of this project, and Michael Wee, for his guidance. We also like to show our appreciation to our friends, who stood by us when we were struggling with the difficulties in this project.

[Pau01]   Paul, Seth. 2001. Microsoft SQL Server 2000: Third-Party Data Mining Providers. White Paper from Microsoft Corporation.

[Tod00]   Bill, Todd. 2000. Introduction to InterBaseExpress. 11[th] Annual Borland Conference, Proceedings. Borland Developers Network, Article ID: 27201

[Tod01]   Todd, Bill. 2001. InterBase: What Sets It Apart. Borland Developers Network, Article ID: 27007

[War99]   Ward, Nick. 1999. Sybase Industry Warehouse Studio: A Technical Overview. White Paper from Sybase, Inc.

APPENDIX

*Due to space constraints, data structures and definitions are highly reduced.

Conventions:
→ means put into
➔ means implies
= means schema definition
TD() means tuple definition

Preliminary Definitions:
[Statement]
[Word]

We define a dataset as,

DataSet = [Table:ΠWord; SQLStatement:Statement| TD(Table$_{i=1}^{\Pi}$); init Table = 0]

And SQLAction >> DataSet, where,

SQLAction = [ ΔDataSet; Success::==Yes|No; Table:ΠWord; SQLStatement:Statement | TD(Table$_{i=1}^{\Pi}$); Table! = ΥDatabase? •(x|x specified by SQLStatement?); SQLStatement!= SQLStatement?; (SQLStatement? ε SQL92)➔ Success!=YES; !(SQLStatement? ε SQL92)➔ Success!=NO]

SQL92 refers to ISO SQL-92 Specification.

The kernel component is defined as,

Kernel = KernelBinder Υ IAP Υ ToolController Υ WizardController
Where
KernelBinder = Υ(KernelTable, KernelAction, DataSet)
IAP = Υ(DataSet, IBSpec, IAPAction, IBInfo!)
ToolController = ToolAction extends [Action?:Word; Success!:Word| (Action!=AddTool)➔ AddTool; (Action!=RemoveTool)➔ RemoveTool; (Action!=Use)➔ AllocTool; (Action!=Unuse)➔ DeallocTool]

KernelTable encapsulates the data structures used by the KernelBinder.
IBSpec encapsulates connection-specific information, which is delivered as IBInfo!

KernelAction = [ Δ(KernelTable^IAP); ActionCode?, DatabaseName?, Path?, TableName?, User?, Password?, SQLStatement?, Success!:Word| ((Path?^User? ε KernelTable)➔ (IAP'$_{(User?,Path?)}$ ← SQLStatement?); !(Path?^User? ε KernelTable)➔ ((Path?, User?)➔ KernelTable(Path', User'))^ (ActionCode?, DatabaseName?, Path?, TableName?, User?, Password?, SQLStatement?)➔ (IAP'$_{(User?,Path?)}$ (ActionCode?, DatabaseName?, Path?, TableName?, User?, Password?, SQLStatement?))); Success! ← IAP'$_{(User?,Path?)}$(Success!); ((CurrentMemory ε IBInfo!) = 0) ➔ (KernelTable' = KernelTable – ({Path?}^{User?}))]

ToolAction = [ ΔTActiveTable; Υ(ToolTable, TActiveTable, AddTool, RemoveTool, DisplayTool, AllocTool, UnallocTool) extends (KernelAction Υ DataSet)

AddTool = [ΔToolTable; Name?, Path?, Initialization?, Description?, Author?, Success!: Word; Version:N| Ε(Name? ε ToolTable Υ User? ε ToolTable) ➔ Success!=NO➔ ("Tool already installed"); Ε(Name? ε ToolTable Υ User? ε ToolTable)^(Version$_{Name}$ > Version?) ➔Success!=NO➔ ("Attempt to install older version"); Ω(Name? ε ToolTable Υ User? ε ToolTable) = 0 ➔((Name?, Version?, Initialization?, Description?, Author?)➔ ToolTable'(Name, Version, Initialization, Description, Author))➔ Success!=YES]

AllocTool = [ ΔTActiveTable; Name?, User?, Success!: Word| (Name? ε ToolTable)^!((Name?^User?) ε TActiveTable) ➔(Name?, User? → TActiveTable'(Name, User)➔ init(Path•Path$_{Name}$ =Name?)^ init(Initialization•Initialization$_{Name}$ =Name?) ➔ Success! = YES; ((Name?^User?) ε TActiveTable) v !(Name? ε ToolTable) ➔ Success!=NO]

TActiveTable holds the tools have are in use.
ToolTable is the master table which holds information for all tool modules.

The Discovery Platform is defined as,

Discoverer = [ΔDModuleTable; Υ(DModuleAdder, DisplayDTable, DModuleRemoverer)]

DModuleAdder = [ΔDModuleTable^(ToolTable ϖ WizardTable); Name?, Path?, Type?, Initialization?, Description?, Author?:Word; Version?:N| !(Name?, Version? = DModuleTable(Name, Version)) → (((Name?, Path?, Type?, Description?) → DModuleTable'(Name, Path, Type, Description)); (Type? = Tool)→ (DModuleTable>>AddTool); (Type?=Wizard)→(DModuleTable>>AddWizard))]

Multi-D Cube Builder is defined as,

MDCB = [ ΔCubeTable; ΔDimensionList; ϒ(DisplayCube, DisplayAllCube, AddDimension, AddCube, RemoveDimension, RemoveCube)]

AddCube = [ΔCubeTable; Name?, Owner?, Description?, Success!:Word; DList:DimensionList| Ω(CubeName ε CubeTable) = 0 →(((Name?, Owner?, Description?)→ CubeName'(Name, Owner, Description))^CubeName'(DList) ← (Name? >> ϒ$_{i=1}^{n}$ AddDimension$_i$)) →(Success! = YES); Ω(CubeName ε CubeTable)= 1→ (Success! = NO)]

AddDimension = [ΔDimensionList; CubeName?, Database?, Table?, Dimension?, Success!: Word; DNumber?:N| Ω(ϒ(CubeName?, Database?, Table?, Dimension?) ε DimensionList) = 0→ ((CubeName?, Database?, Table?, DNumber?, Dimension?)→ DimensionList'(CubeName, Database, Table, DNumber, Dimension))→ Success!=YES; Ω(ϒ(CubeName?, Database?, Table?, Dimension?) ε DimensionList) = 1 → Success!=NO]